\documentclass[epj]{svjour}

\usepackage{graphics}

\begin{document}

\title{Coherent Thomson backscattering
from laser-driven relativistic ultra-thin electron layers}
\author{J. Meyer-ter-Vehn and Hui-Chun Wu}
\institute{Max-Planck-Institute for Quantum Optics, D 85748 Garching, Germany}
\date{Received: date / Revised version:}
%\subtitle{Do you have a subtitle?\\ If so, write it here}
% etc
% \thanks is optional - remove next line if not needed
%\thanks{\emph{Present address:} Insert the address here if needed}%
%}                     % Do not remove
%
%\offprints{}          % Insert a name or remove this line
%
% The correct dates will be entered by Springer
%
\abstract{ The generation of laser-driven dense relativistic
electron layers from ultra-thin foils and their use for
coherent Thomson backscattering is discussed, applying analytic
theory and one-dimensional particle-in-cell simulation. The blow-out
regime is explored in which all foil electrons are separated from
ions by direct laser action. The electrons follow the light wave
close to its leading front. Single electron solutions are applied to
initial acceleration, phase switching, and second-stage boosting.
Coherently reflected light shows Doppler-shifted spectra, chirped
over several octaves. The Doppler shift is found $\propto
\gamma_x^2=1/(1-\beta_x^2)$, where $\beta_x$ is the electron
velocity component in normal direction of the electron layer which
is also the direction of the driving laser pulse. Due to transverse
electron momentum $p_y$, the Doppler shift by
$4\gamma_x^2=4\gamma^2/(1+(p_y/mc)^2)\approx 2\gamma$
is significantly smaller than full shift of $4\gamma^2$.
Methods to turn $p_y\rightarrow 0$ and to recover the
full Doppler shift are proposed and verified by 1D-PIC simulation.
These methods open new ways to design intense single attosecond
pulses.  }
\authorrunning{ J. Meyer-ter-Vehn and H.-C. Wu}
\titlerunning{Coherent Thomson backscattering from relativistic ultra-thin electron layers}
\maketitle

\begin{figure*}[t]
\label{fig1}
\centering\resizebox{0.75\textwidth}{!}{\includegraphics{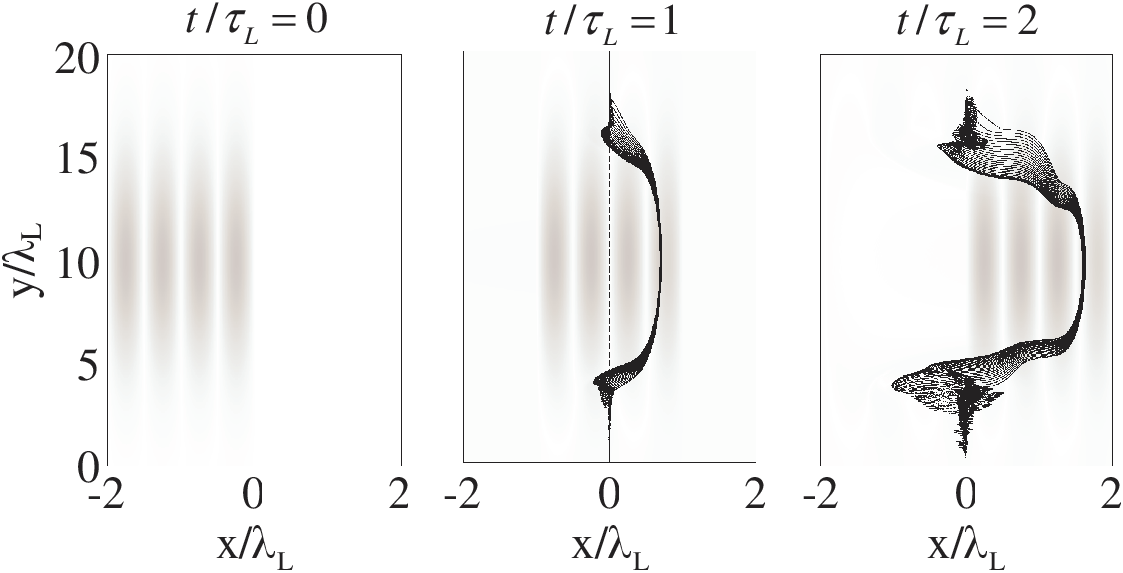}}
\caption{ Irradiation of a graphene layer with a two-cycle laser
pulse of $a_{0}=5$ (2D-PIC simulation). Interfaces of electron
layers are plotted in the plane of laser polarization before as well
as one and two laser cycles after the light front has touched the
foil. The simulated laser intensity is shown in shaded grey;
the broken line shows the immobile ion layer left behind.}
\end{figure*}

\section{Introduction}

\label{intro} There is a quest for high-quality x-ray sources in many fields
of science \cite{Salam}. Coherent VUV- and X-ray sources now become
available at XFEL facilities \cite{Ackermann}. Also the generation of high
harmonics from optical laser pulses has provided us with very useful \
VUV-sources recently \cite{Seres}.

In this paper, we deal with Thomson scattering from relativistic electrons.
Incoherent Thomson scattering converts photons of incident frequency $\omega_L$
to frequencies $\omega =\gamma ^{2}(1-\beta \cos \theta )^{2}\omega_L$ due to
relativistic Doppler shift \cite{Einstein}; here $\beta =v/c$ is electron
velocity normalized to light velocity $c$, $\theta $ is the scattering angle,
and $\gamma =1/\sqrt{1-\beta^{2}}$ the usual relativistic factor. Thomson
scattering has been measured from high-$\gamma $ electrons at conventional
accelerator facilities \cite{Esarey} and recently also from laser
generated electron beams \cite{Schwoerer}. In these cases, light was
backscattered incoherently.

Coherently backscattered photons can be obtained from electron
layers provided they are dense enough so that several
backscattering electrons reside in the volume $\lambda_{R}^{3}$ ,
where $\lambda _{R}$ is the wavelength of the light in the rest
frame of the electron layer. When moving close to velocity of
light, we refer to these layers as\textit{\ relativistic mirrors} with $%
\gamma $-factors $\gamma _{m}$ \cite{Bulanov}. Such mirrors have
been created recently in form of nonlinear plasma waves near the
point of wave-breaking when they develop diverging density spikes;
values of $\gamma _{m}=4-5$ could be deduced from the backscattered
spectrum \cite{Pirozhkov}.

Relativistic mirrors are also formed when irradiating solid surfaces
with laser amplitudes $a_{0}=eA_{L0}/mc^{2}>1$
\cite{Lichters,Gordienko,Tsakiris}; here $e$ and $m $ are electron
charge and mass, respectively, and $A_{L0}$ is the amplitude of the
vector potential of the light wave. For $a_{0}>1,$ the solid surface
turns into dense plasma, and electrons move close to velocity of
light. For linear polarized light, the ponderomotive force has a
component oscillating at twice the laser frequency $\omega _{L}$.
This leads to an oscillating mirror front and to high harmonics in
the reflected light. These surface harmonics have been observed
recently \cite{Dromey}. The results show harmonic orders up to 3200
corresponding to 3.7 keV photons. They are in agreement with theory
\cite{Baeva}.

In the present paper, we consider laser irradiation of ultra-thin
(nm) foils \cite{Vshivkov,Pirozhkov2006,Mikhailova,Rykovanov} at
intensities high enough to separate all electrons from ions. The
electrons then move with the front of the laser pulse, forming a
dense relativistic electron layer. This regime has been investigated
recently by Kulagin et al. \cite{Kulagin}. Here it is studied for
Thomson backscattering of a counter-propagating probe pulse.

For separating electrons from ions, the laser field $a_{0}=E_{L0}$
has to be larger than the total electric field $\varepsilon_0$
that builds up due to charge separation.
Here both the laser field $E_{L0}=k_{L}A_{L0}$ with $k_{L}=\omega _{L}/c$
and the longitudinal field $\varepsilon_0$ are taken in units
of $E_{0}=mc\omega _{L}/e$. For a planar foil with electron
density $n_{e}$ and thickness $d$, we find $\varepsilon_0=Nk_{L}d$.
Here \ $N=n_{e}/n_{crit}=\omega_{p}^{2}/\omega _{L}^{2}$ is the
density normalized to the critical density $n_{crit}$ and can be
conveniently expressed by the ratio of plasma frequency $\omega
_{p}=(4\pi e^{2}n_{e}/m)^{1/2}$ and laser frequency $\omega _{L}$.
The condition for complete electron blow-out is
\begin{equation}\label{eq1}
a_{0}\gg Nk_{L}d.
\end{equation}
As a reference foil material, we consider in the following graphene which
consists of a mono-atomic layer of carbon atoms arranged in regular hexagons
with a side length of $s=0.141$ nm. Having an area of
$A=(3\sqrt{3}/2)s^{2}=5.2\times 10^{-16}$cm$^{2}$ and
a total of 12 electrons attached to each hexagon,
the electron areal density is $n_{e}d=12/A=2.3\times 10^{16}$cm$^{-2}$,
and one finds $\varepsilon _{0}=Nk_{L}d$ $\simeq 1$.
Graphene layers have been successfully produced in pieces up to
$100\times 100$ $\mu $m$^{2}$ \cite{Geim}. This is sufficient in principle
for experiments with existing laser pulses in the range of $a_{0}\approx 3-10$,
corresponding to intensities of \ $10^{19}-10^{20}$W/cm$^{2}$.
It is interesting to notice that an area of $10\times 10$ $\mu $m$^{2}$
contains a total charge of about 1 nC, comparable and exceeding values
in other recent experiments using laser wakefield acceleration and gas targets.
Alternatively, multi-layer carbon foils with thicknesses of a few nm
are available for experiments, but require higher laser intensities for
complete electron blow-out.

For illustration, a two-dimensional (2D) particle-in-cell (PIC) simulation
of a graphene layer irradiated by a laser pulse is shown in Fig. 1.
Electron density is plotted for three different times, and
also the driving laser pulse is shown. Here we have chosen
a two-cycle pulse $E(\tau)/E_{0}=a_{0}\sin \tau $ with
$\ 0\leq \tau \leq 4\pi $ and $a_{0}=5$.
At the corresponding intensity of $3\times 10^{19}$W/cm$^{2}$,
all electrons within the focal spot are taken as free initially,
i.e. ionization is not treated explicitly. It is seen that the
electrons of the irradiated spot have separated from the ions and are moving
with the laser pulse, surfing as a dense layer near the pulse front. One
should notice that these ultra-thin foils are transparent to the light, even
though they have densities far above the critical density.
This is because they are much thinner than the skin depth
$\lambda _{skin}$ $\approx c/\omega _{p}\approx \lambda _{L}/10$.
The light is therefore penetrating through the foil,
and each electron is accelerated according to the local laser field.
This means that we can use the analytic solution for relativistic electron
motion in a plane laser field as an approximation,
supplemented by the space charge electric field $E_{x}$.
It is constant in time for each sub-layer of the foil,
as long as these sub-layers keep their relative order and do not
overtake each other. This option for analytical treatment is an important
feature of ultra-thin foil acceleration. It is quite different from
usual relativistic laser plasma interactions, e.g. encountered
in laser wake field acceleration, where often complicated
plasma fields dominate.

Here it is direct interaction with the external laser fields
that accelerates the electrons. As we shall discuss in section 2,
electrons gain momentum up to a maximum of
$\mathbf{p}_{\max }=(p_{x},p_{y})=(2a_{0}^{2}mc,2a_{0}mc)$
and energy up to $W_{\max}=(\gamma -1)mc^{2}=2a_{0}^{2}mc^{2}$
in the first half-wave of the light pulse, at least those at the layer front
where $E_x=0$. For the laser pulse $a_{0}=5$ used in Fig.1,
the electron energy rises up to $W_{\max }\approx 25$ MeV
according to these estimates,
but then drops again to zero in the second half of the laser wave.
This behavior reflects the Lawson-Woodward theorem
\cite{Lawson-Woodward}, stating that there is no net energy transfer
to charged particles by light waves in vacuum.
Having in mind generation of X-rays by Thomson scattering,
this is acceptable, because there is plenty of time for
back-scattering of a probe pulse by the dense electron sheath
while surfing on the wave. Actually, the time available is approximately
$t=(3a_{0}^{2}/8)\tau _{L}$ which is about 10 laser periods $\tau _{L}$ for
$a_{0}=5$. For $\gamma \approx 2a_{0}^{2}=50$, one might be tempted to
expect coherent photons Doppler-shifted by a factor  $4\gamma ^{2}=10^{4}$.
Unfortunately, this is not the case because the effective $\gamma$-factor
at which the layer is moving in normal $x$-direction is
\begin{equation}
\gamma _{x}=1/\sqrt{1-\beta _{x}^{2}}=\gamma /\sqrt{1+(p_{y}/mc)^{2}}\approx
\sqrt{\gamma /2},
\end{equation}
and the frequency is shifted by a factor $4\gamma _{x}^{2}\approx 2\gamma $
rather than $4\gamma ^{2}$. The reason for this is inherent to direct laser
acceleration which necessarily involves large transverse momenta
$p_{y} \propto a_{0}$ for large $\gamma \propto p_{x} \propto a_{0}^{2}$.
Electrons move under an angle $\theta \approx p_{y}/p_{x}\propto 1/a_{0}$
relative to the laser direction. For incoherent Thomson scattering,
the reflected signal is emitted close to the direction of the individual
high-$\gamma $ electrons and show the angle $\theta $. In the case of
coherent scattering from a planar layer, however, this main component is suppressed
by destructive interference and only the component scattered into the
normal direction of the sheet will survive. Since incoherent scattering is proportional
to the number $N_{s}$ of scatterers, while coherent scattering is $\propto N_{s}^{2}$,
the latter is bound to win as the layer becomes dense enough.
In practice, one may encounter cases intermediate between incoherent and
coherent scattering and observe the change in direction, frequency, and
intensity.

The present paper is intended to explore basic features of electron sheet acceleration
and back-scattering in 1D geometry. The analysis is restricted to schematic
laser pulses of single-cycle and half-cycle form. In section 2, we review
the analytic solution of single-electron motion in a planar light wave and
discuss three relevant situations: (1) acceleration of an electron initially
at rest, (2) collision of a relativistic electron with a counter-propagating
wave, and (3) boost acceleration of an electron entering the light wave already at
relativistic energy. Results of cases (1) and (2) are then compared
with 1D-PIC simulations in section 3, including back-scattering
of a counter-propagating probe pulse.
The coherent reflectivity of an ultra-thin relativistic layer is derived
analytically in a companion paper, referred to as paper II \cite{Wu}.
The full analytical description of the driven electron layer,
including self-radiation and space charge fields,
was developed by Kulagin \cite{Kulagin2007}; it will be applied
to the present reference case in another companion paper \cite{Wen}.

\section{Single electron in plane laser wave}

\subsection{Equations of motion}

Here we review the analytical treatment of a relativistic electron
interacting with a plane light wave in vacuum, to the extent it is needed in the
following discussion \cite{MtV}. The linear polarized light wave propagates
in $x$-direction and has electric and magnetic field components $E_{y}$ and $B_{z}$,
respectively. In addition, we allow for a longitudinal $E_{x}$ field.
The equation of motion then read

\begin{eqnarray}
\frac{dp_{x}}{dt} &=&-E_{x}-\beta _{y}B_{z},  \nonumber \\
\frac{dp_{y}}{dt} &=&-E_{y}+\beta _{x}B_{z}, \nonumber\\
\gamma ^{2} &=&1+p_{x}^{2}+p_{y}^{2},  \nonumber \\
\frac{d\gamma }{dt} &=&-\beta _{x}E_{x}-\beta _{y}E_{y}, \\
\frac{dx}{dt} &=&\beta _{x}=p_{x}/\gamma , \nonumber\\
\frac{dy}{dt} &=&\beta _{y}=p_{y}/\gamma \nonumber.
\end{eqnarray}
Here we use dimensionless variables corresponding to dimensional ones
according to
\begin{eqnarray*}
(t,x,y)\,&\widehat{=}&\,(\omega_Lt,k_Lx,k_Ly), \\
(p_{x},p_{y})\,&\widehat{=}&\,(p_{x}/mc,p_{y}/mc), \\
(E_{x},E_{y},B_{z})\,&\widehat{=}&\,(E_{x},E_{y},B_{z})e/mc\omega_L,
\end{eqnarray*}%
where $\omega_L$ and $k_L$ are circular frequency and wave-number of
the light wave, $e$ and $m$ are charge and mass of the electron, and $c$ is
the velocity of light. For a plane wave moving in vacuum in $x$-direction
with dispersion relation $\omega_L=ck_L$, the laser fields satisfy
$E_{y}(\tau )=B_{z}(\tau )$ with propagation coordinate $\tau =t-x$. For an
electron moving along $x(t)$ , we have $d\tau /dt=1-\beta _{x}$. With these
relations, the equations of motion can be written with $\tau $ as
independent coordinate in the form
\begin{eqnarray}
&&\frac{d}{d\tau }(\gamma -p_{x}) =E_{x},\label{kappa-eq} \\
&&\frac{dp_{y}}{d\tau } = -E_{y}\label{py-eq}.
\end{eqnarray}
Integration gives the functions $\kappa (\tau )=\gamma -p_{x}$ and
$p_{y}(\tau ),$ and, recalling $\gamma ^{2}=1+p_{x}^{2}+p_{y}^{2}$ , we obtain
\begin{eqnarray}
\gamma (\tau ) &=&\frac{1+p_{y}^{2}}{2\kappa }+\frac{\kappa }{2}, \label{gamma-eq}\\
p_{x}(\tau ) &=&\frac{1+p_{y}^{2}}{2\kappa }-\frac{\kappa }{2}
\label{px-eq}.
\end{eqnarray}
Making use of \ $\kappa =\gamma -p_{x}=\gamma (1-\beta _{x})$ , the electron
trajectories $x(\tau )$, $y(\tau )$ follow from
\begin{eqnarray}
dx/d\tau &=&p_{x}/\kappa , \label{x-eq}\\
dy/d\tau &=&p_{y}/\kappa ,\label{y-eq}
\end{eqnarray}
in parametric form with time \ $t(\tau )=\tau +x(\tau )$ .

Let us consider solutions for a single cycle light wave of the form
$E_{y}=a_{0}\sin \tau $, limited to the interval $0\leq \tau \leq 2\pi$
and with no longitudinal field, $E_{x}=0$ . In this case, the solutions of
Eqs. (\ref{kappa-eq}) and (\ref{py-eq}) and  are
\begin{eqnarray*}
p_{y}(\tau ) &=&p_{y0}-a_{0}(1-\cos \tau ), \\
\kappa (\tau ) &=&\gamma (\tau )-p_{x}(\tau )=\kappa _{0},
\end{eqnarray*}%
where $\kappa _{0}$ and $p_{y0}$ are integration constants.
We now discuss three important cases.

%%%%%%%%%%%%%%%%%%%%%%%%%%%%%%%%%%%%%%%%
\begin{figure*}[t]
\label{fig2}
\centering\resizebox{0.50\textwidth}{!} {\includegraphics{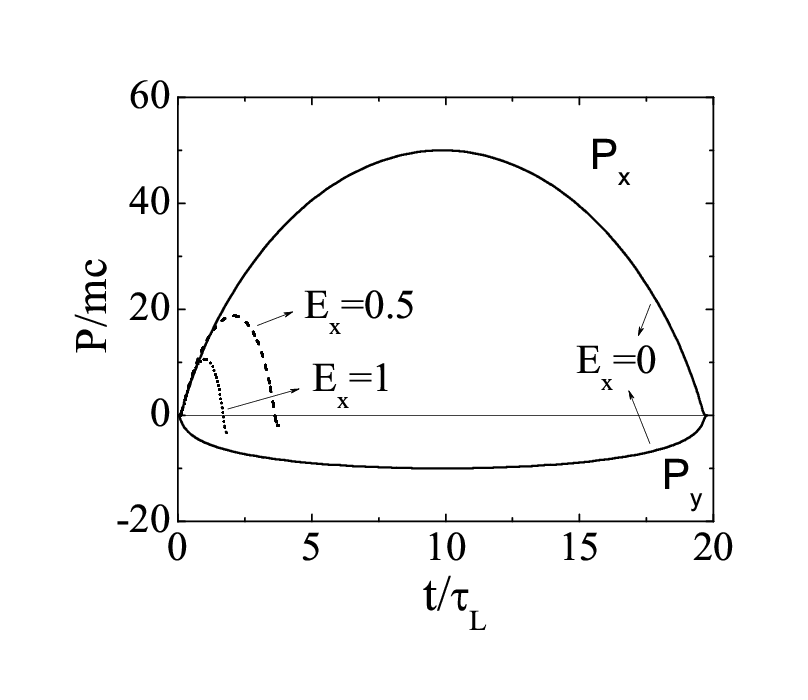}}
\caption{Analytic solution of longitudinal momentum $p_x$
and transverse momentum $p_y$ as function of time for an electron
initially at rest and driven by a single-cycle laser pulse with
$a_{0}=5$. Electrons at different depth in a layer are considered
experiencing different constant longitudinal electric fields
$E_x$; $E_x=0$ corresponds to the right-hand surface
of the layer, $E_x=0.5$ to the middle, and $E_x=1$ to the
left-hand surface facing the ions of the graphene layer. For units see text.}
\end{figure*}
%%%%%%%%%%%%%%%%%%%%%%%%%%%%%%%%%%%%%%%%%%

\subsection{Acceleration of electron initially at rest:
$\protect\kappa_{0}=1$}

An electron initially at rest with $p_{x0}=p_{y0}=0$ and $\gamma _{0}=1$
corresponds to $\kappa _{0}=1$. In this case the explicit solution is
\begin{eqnarray*}
p_{y}(\tau ) &=&-a_{0}(1-\cos \tau ), \\
p_{x}(\tau ) &=&\frac{a_{0}^{2}}{2}(1-\cos \tau )^{2}, \\
\gamma (\tau ) &=&1+\frac{a_{0}^{2}}{2}(1-\cos \tau )^{2}, \\
y(\tau ) &=&a_{0}\int_{0}^{\tau}{(\cos \tau ^{\prime }-1) d\tau^{\prime }}%
=-a_{0}(\tau -\sin \tau ), \\
x(\tau ) &=&\frac{a_{0}^{2}}{2}\int_{0}^{\tau}{(\cos \tau ^{\prime
}-1)^{2}d\tau ^{\prime }}\\
&=&\frac{a_{0}^{2}}{2}(\frac{3}{2}\tau -2\sin \tau +
\frac{1}{4}\sin 2\tau ), \\
t(\tau ) &=&\tau +x(\tau ).
\end{eqnarray*}
For $\tau =\pi $ , the electron reaches the maximum energy of
$\gamma_{1}=1+2a_{0}^{2}$ at time $t_{1}=\pi +x(\pi )=\pi (1+3a_{0}^{2}/4)$,
and momenta and positions then are $p_{x1}=2a_{0}^{2}$, $p_{y1}=-2a_{0}$,
$x_{1}=(3\pi /4)a_{0}^{2}$, and $y_{1}=-\pi a_{0}$. Momenta $p_x$ and $p_y$
are plotted in Fig. 2 as function of time. It is seen that they grow
while interacting with the first half of the driving wave ($0\leq\tau\leq =\pi $);
afterwards, in the second half, the particle decelerates again and
comes to rest at $\tau =2\pi $. There is no net energy transfer
between a particle and the light wave, unless additional fields interfere.
This corresponds to the Lawson-Woodward theorem \cite{Lawson-Woodward}.
For a clear interpretation of this theorem,
one should refer to Esarey et al. \cite{Esarey 1995}.
In order to continue acceleration also in the second half,
one has to shift the transverse momentum $p_{y}$ at $\tau =\pi$
from negative to positive values. In principle, this is possible
by interaction with an additional counter-propagating laser pulse
and will be discussed next as a second interesting solution
of the basic equations.

\subsection{Collision with relativistic electron:
$\protect\kappa _{0}>>1$}

Let us now consider an electron, moving opposite to the direction of the
light pulse and having initial momenta $p_{x1}=-2a_{0}^{2}$,
$p_{y1}=+2a_{0}$, and energy $\gamma _{1}$ $=1+2a_{0}^{2}$.
This is equivalent to the case considered before at the time,
when the electron has peak energy after interaction with the first half-wave.
At this moment, it is hit head-on by the second pulse running in opposite direction.
We call this second pulse the \textit{switch pulse},
because it almost instantaneously changes $p_{y}$
(and thereby the phase between electron and wave),
but leaves $p_{x}$ and $\gamma$ almost unchanged.
This type of interaction corresponds to
$\kappa _{1}=\gamma_{1}-p_{x1}=1+4a_{0}^{2} \gg 1$ for $a_{0}>1$.
Placing the electron again at the origin $x_{1}=y_{1}=0$ at time $\tau =0$,
we obtain the solution
\begin{eqnarray*}
p_{y}(\tau ) &=&a_{0}(1+\cos \tau ), \\
p_{x}(\tau ) &=&\frac{1+a_{0}^{2}(1+\cos \tau )^{2}}{2(1+4a_{0}^{2})}-
\frac{1+4a_{0}^{2}}{2}, \\
\gamma (\tau ) &=&\frac{1+a_{0}^{2}(1+\cos \tau )^{2}}{2(1+4a_{0}^{2})}+
\frac{1+4a_{0}^{2}}{2}, \\
y(\tau ) &=&a_{0}(\tau +\sin \tau )/(1+4a_0^2), \\
x(\tau ) &=&\frac{(1+3a_{0}^{2}/2)\tau +2a_{0}^{2}\sin \tau
+(a_{0}^{2}/4)\sin 2\tau )}{2(1+4a_{0}^{2})^{2}}-\frac{\tau }{2}, \\
t(\tau ) &=&\tau +x(\tau ),
\end{eqnarray*}
valid for $0\leq \tau \leq 2\pi $.
Apparently, the dynamics are very different from the first case. Only the
transverse motion scales $\propto a_{0}$ as before,
while all changes of longitudinal quantities are small of order $1$ .
During the first half cycle $(0\leq \tau \leq \pi )$,
transverse momentum $p_{y}$ turns from $2a_{0}$ to $0$ ,
while energy and longitudinal momentum change as
$\Delta \gamma =\Delta p_{x}=2a_{0}^{2}/(1+4a_{0}^{2})\approx 1/2$.
This takes place in the short time and space interval
$\Delta x\approx \Delta t\approx \pi /2$\,
corresponding to a quarter cycle.
A half-cycle laser pulse of same strength as the drive pulse,
but running in opposite direction, is therefore sufficient
to switch the phase of the electron from deceleration back to acceleration.
In case the switch pulse extends over a full cycle, the situation
of $p_y\approx 0$ can be realized only for short moment.
The action of both half- and full-cycle pulses will be compared
with 1D-PIC simulation in section 3.

\subsection{Boosting energy of a relativistic electron:
$\left\vert \protect\kappa _{0}\right\vert <<1$}.

A third, very instructive solution of Eqs. (\ref{intro}) to (\ref{y-eq}) is obtained
for $\left\vert \kappa _{0}\right\vert <<1$.
Here we start with an electron that has already been accelerated
by the first half-wave of case 1 and has interacted
with the half-wave switch pulse of case 2.
Initial momenta and energy now are
\begin{eqnarray*}
p_{y2} &=&0, \\
p_{x2} &=&-\frac{1}{2(1+4a_{0}^{2})}+\frac{(1+4a_{0}^{2})}{2}, \\
\gamma _{2} &=&\frac{1}{2(1+4a_{0}^{2})}+\frac{(1+4a_{0}^{2})}{2}, \\
\kappa _{2} &=&\gamma _{2}-p_{x2}=\frac{1}{(1+4a_{0}^{2})}.
\end{eqnarray*}%
Entering with these values into the negative half-wave of
$E_{y}=a_{0}\sin \tau $ at time $\tau =\pi $ , the electron moves in the time
interval $\pi \leq \tau \leq 2\pi $ according to
\begin{eqnarray*}
p_{y}(\tau ) &=&a_{0}(1+\cos \tau ), \\
p_{x}(\tau ) &=&\frac{(1+4a_{0}^{2})(1+a_{0}^{2}(1+\cos \tau )^{2})}{2}-%
\frac{1}{2(1+4a_{0}^{2})}, \\
\gamma (\tau ) &=&\frac{(1+4a_{0}^{2})(1+a_{0}^{2}(1+\cos \tau )^{2})}{2}+%
\frac{1}{2(1+4a_{0}^{2})}.
\end{eqnarray*}%
For $\tau =2\pi $ , we find $\ \gamma _{2}\approx p_{x2}\approx
8a_{0}^{4}$ , and the leading orders of the shifts in x and y direction are
\ $\Delta x\approx 12a_{0}^{6}\pi $\ \ and \ $\Delta y\approx 4a_{0}^{3}\pi$, always supposed that $a_{0}>>1$. Choosing a laser pulse with $a_{0}=5$ and
wavelength \ $\lambda_L=0.8$ $\mu $m, the single-cycle sin-pulse
accelerates the electron to an energy of \ $W=\gamma mc^{2}\approx 2.5$ GeV.
For this it needs a distance of \ $\Delta x\approx 6a_{0}^{6}\lambda
_{L}\approx 7.5$ cm and a time interval of \ $\Delta t\approx \Delta
x/c\approx 250$ ps.

We show these results here as an interesting application of the
one-dimensional relativistic equations of motion.
Of course, the example is difficult to realize experimentally for several reasons.
It is not clear how to produce and transport the half-cycle switch pulse.
%and, also, diffraction will limit the validity range of this one-dimensional solution.
%The half-cycle pulses
It has a spectrum with a non-zero dc component and
cannot be transported over a longer distance due to strong diffraction.
An alternative method to switch the transverse momentum $p_{y}\rightarrow 0$
would be a local static $B_{z}$ field. This would require
$\left\vert p_{y}\right\vert =a_{z}k_LD=2a_{0}$,
where $D$ is the thickness of the magnetic layer and $a_{z}=B_{z}/B_{0}$
with $B_{0}=mc\omega_L/e\approx 10^{4}$ T$\cdot \mu $m/$\lambda _{L}$.
Since the switch layer should not be thicker than a few $\lambda _{L}$,
we would need a magnetic field in the order of
\ $B_{z}\approx B_{0}a_{0}/10\approx a_{0}$ kT,
which is quite demanding, but may be not impossible.

\subsection{Charge separation and longitudinal fields $E_{x}$}

So far, we have considered solutions for single electrons only.
In the case of a solid-density layer and the expulsion of all electrons,
large longitudinal fields $E_{x}$ will emerge due to space charge
and will modify the solutions given above.
%Here we briefly address this problem, but leave
%a detailed analytical treatment to a separate publication.
For a plane foil of thickness $d$ and electron density $n_{e}$,
the electric field between the layers of the expelled electrons and
the ions left behind is given by
\begin{equation}
\varepsilon _{0}=Nk_Ld,
\end{equation}
where $\varepsilon _{0}$ is the electric field in units of $E_{0}=mc\omega_L/e$ and
\begin{equation}
N=n_{e}/n_{crit}=\omega _{p}^{2}/\omega_L^{2}
\end{equation}
is the electron density, normalized to the critical density
$n_{crit}=\pi mc^{2}/(e\lambda_L)^{2}\simeq 1.1\times 10^{21}$cm$^{-3}$/$\lambda_L^{2}[\mu m]$.
Electrons, initially located at $x=x_{0}$ inside the electron layer, experience the electric field
\begin{equation}
E_{x}=\varepsilon _{0}(1-x_{0}/d),
\end{equation}
which is largest for $x_{0}=0$ on the laser-irradiated front side of the foil
and vanishes at $x_{0}=d$ on the rear side. Neglecting the initial time period of
electron expulsion from the foil, which is short for ultra-thin foils,
the field $E_{x}$ is constant in time. This holds as long as the total charges
to the left and to the right of $x_{0}$ stay the same,
i.e. sub-layers do not change their relative order.
We can then integrate $d\kappa /d\tau =E_{x}$ and find
\begin{equation}
\kappa (\tau )=1+\varepsilon _{0}(1-x_{0}/d)\tau.
\end{equation}
The full solution is obtained by integrating Eqs. (\ref{kappa-eq}) - (\ref{y-eq}).
In Fig. 2 we have also plotted the curves $p_{x}(t)$
including  the effect of $E_x$ for the surface on the laser side ($x_{0}/d=0$)
and the middle interface ($x_{0}/d=0.5$) of the electron layer.
For graphene the fields are $E_x =1$ and $E_x =0.5$, respectively.
It is seen that the drag-back of the ions is very strong
when compared to the leading front of the layer at $x_{0}/d=1$.
Only at early times the layer moves as a whole,
but then electrons facing the ions lag behind and
even turn around and fall back to the ion layer.

It should be noticed that the evolution of the electron layers
is still more complex. Almost quantitative agreement with
1D-PIC simulations is obtained \cite{Wen} when applying
the full analytical model of Kulagin \cite{Kulagin2007}.

%%%%%%%%%%%%%%%%%%%%%%%%%%%%%%%%%%%%%%%%%%%%%%%%%%%%
\begin{figure*}[tbph]
\label{fig3} \centering\resizebox{0.75\textwidth}{!}
{\includegraphics{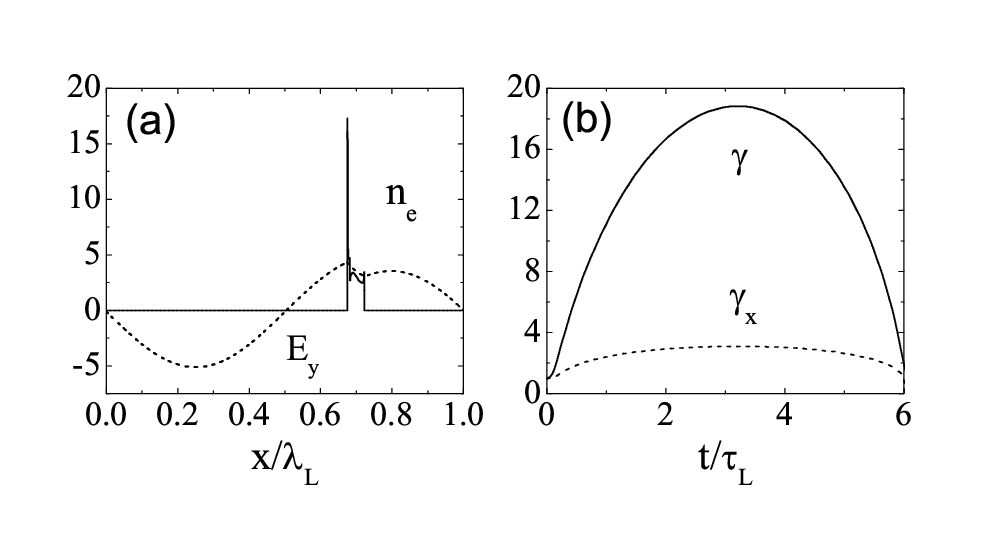}} \caption{Results of 1D-PIC simulation:
(a) Snapshot of laser field $E_y$ and electron density $n_e$ after
one laser cycle when the single-cycle pulse has just passed the
graphene layer located at $x=0$. Electrons are seen as a thin layer
surfing on the first half-cycle and marked by sharp density spikes
at the surface facing the drive laser. (b) Values of $\gamma$ and
$\gamma_x$ of electrons, located at the right surface of the layer,
are plotted as function of $t$.}
\end{figure*}
%%%%%%%%%%%%%%%%%%%%%%%%%%%%%%%%%%%%%%%%%%%%%%%%%%%%%%%%

%%%%%%%%%%%%%%%%%%%%%%%%%%%%%%%%%%%%%%%%%%%%%%%%%%%%%%%%
\begin{figure*}[tbph]
\label{fig4} \centering\resizebox{0.75\textwidth}{!}
{\includegraphics{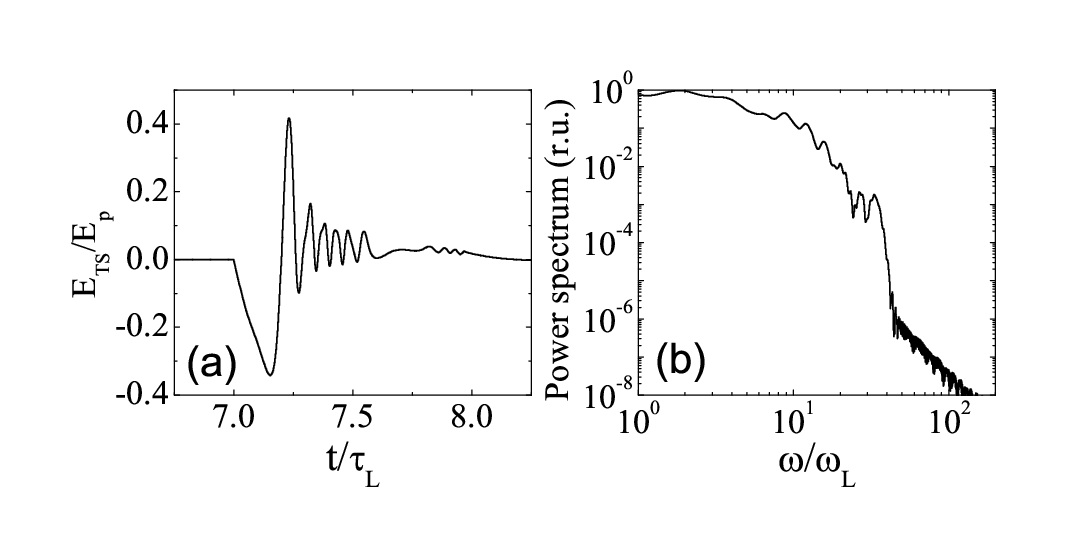}} \caption{ Thomson backscattering of a
semi-infinite probe pulse by the electron layer shown in Fig. 3. The
weak probe pulse with $a_{p}=0.001$ is polarized perpendicular to
the drive pulse, and its sharp front reaches the foil at $t=0$. (a)
Signal versus time as seen by an observer at $7 \lambda_L $ behind
foil, (b) corresponding spectrum, which shows a sharp cutoff at
$\omega/\omega_L\approx 4\gamma_x^2\approx 36$, well corresponding
to the maximum value of $\gamma_x\approx 3$ in Fig.3b.}
\end{figure*}
%%%%%%%%%%%%%%%%%%%%%%%%%%%%%%%%%%%%%%%%%%%%%%%%%%%%%%%%%%%%%%

%%%%%%%%%%%%%%%%%%%%%%%%%%%%%%%%%%%%%%%%%%%%
\begin{figure*}[t]
\label{fig5} \centering\resizebox{0.75\textwidth}{!}
{\includegraphics{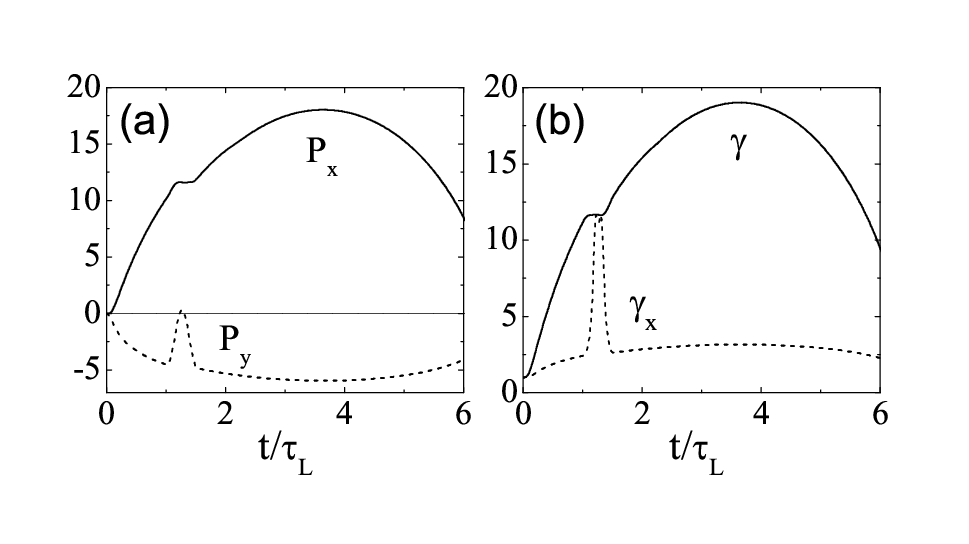}} \caption{Same as the case of Fig. 3,
but applying in addition a full-cycle switch pulse that hits the
accelerated electron layer approximately at $t\approx \tau_L$. The
plots show (a) momenta and (b) $\gamma$ and $\gamma_x$ as function
of time. The switch pulse was chosen such that transverse momentum
$p_y \rightarrow 0$ for a short moment and, correspondingly,
$\gamma_x \rightarrow \gamma$.}
\end{figure*}
%%%%%%%%%%%%%%%%%%%%%%%%%%%%%%%%%%%%%%%%%%%%%

%%%%%%%%%%%%%%%%%%%%%%%%%%%%%%%%%%%%%%%%%%%%%
\begin{figure*}[t]
\label{fig6}
\centering\resizebox{1.\textwidth}{!}{\includegraphics{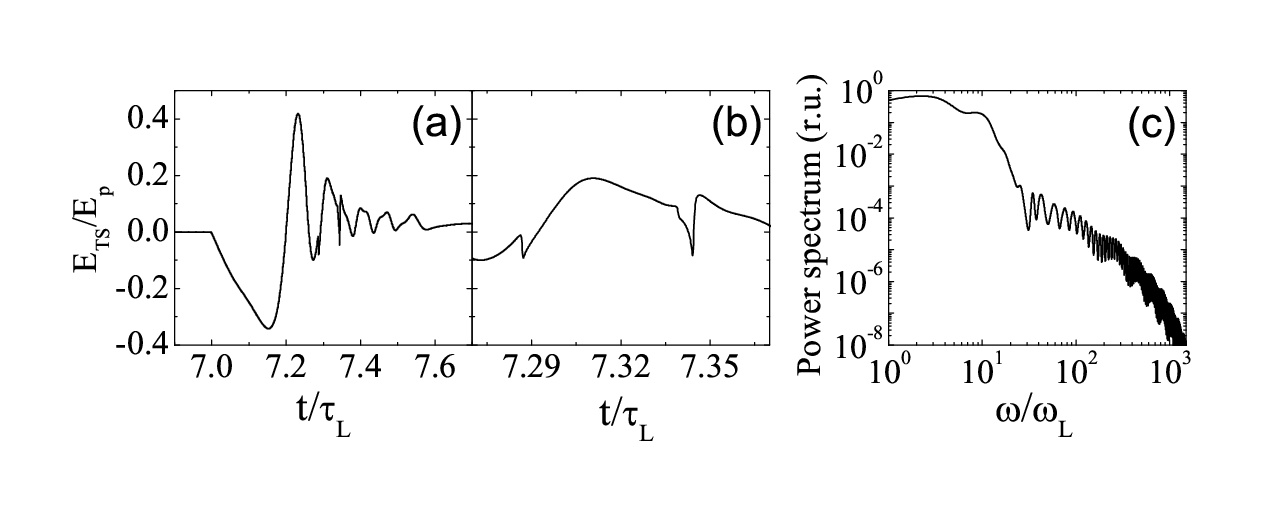}}
\caption{(a,b) Thomson signal and (c) spectrum of probe pulse
corresponding to the case of Fig. 5 with full-cycle switch pulse.
Notice the two conspicuous spikes in the backscattered signal,
zoomed in (b). They generate the high-frequency part in the spectrum
with power-law decay up to $\omega/\omega_L\approx 500$ and a
shoulder extending up to $\omega/\omega_L\approx 2000$. See text for
origin of this behavior. }
\end{figure*}
%%%%%%%%%%%%%%%%%%%%%%%%%%%%%%%%%%%%%%%%%%%%%%%

\section{Simulation}

\subsection{Acceleration of ultra-thin electron layer}

We have performed one-dimensional particle-in-cell (1D-PIC) simulations to
investigate thin layer acceleration and Thomson back-scattering.
For illustration, we choose as before a single-cycle laser pulse
$E_{y}=a_{0}\sin \tau $ with $0\leq \tau \leq 2\pi $, $a_{0}=5$,
and a layer with $\varepsilon _{0}=1$, corresponding to graphene.
A snapshot of electron sheet acceleration is shown in Fig. 3a and the temporal
evolution of $\gamma (t)$ and $\gamma _{x}(t)=\gamma /(1+p_{y}^{2})^{1/2}$ in
Fig. 3b. All electrons are blown out of the foil. The electron density
distribution is plotted here three laser cycles after the laser front has first
interacted with the foil. The thickness of the electron layer is set to
$d/\lambda_L=0.001$ initially and expands to $\Delta x/\lambda_L\approx 0.06$.
The density profile shows a pronounced peak at the surface facing the drive laser.
It surfs on the laser wave and has substantially depleted the front part of the
wave due to energy transfer to electrons. The electrons move close to
velocity of light, but lag somewhat behind the light front at $x/\lambda_L=3$.
Electron $\gamma$-values corresponding to the density spike are seen to
rise to a maximum of $\gamma \approx 19$ and then fall back to unity. Though
this behavior is qualitatively similar to that of the analytical solution
displayed in Fig. 2, the peak value is far less than the analytical maximum of
$\gamma=2a_0^2=50$. The deviation is attributed to modifications in the radiation field
due to energy depletion and self-radiation of the layer. Both effects are
not taken into account in the present analytical treatment.

\subsection{Thomson backscattering}

We now add a weak probe pulse to the simulation above. It propagates
antiparallel to the drive pulse and is polarized orthogonal to it so that
the fields of probe and drive pulse can be clearly separated in the simulation.
The probe pulse is sinusoidal and semi-infinite with a sharp front,
which first hits the foil at $t=0$, simultaneously with the drive pulse.
The back-scattered signal is shown in Fig. 4a, as detected by an observer
located at $x/\lambda_L=7$. The scattered pulse is strongly
chirped due to the acceleration of the electron layer.
Notice that it is also strongly compressed. It vanishes after $t/\tau_L\approx 8$
due to density decay of the layer. The corresponding spectrum in Fig.4b shows a
broad plateau with a sharp cut-off at $\omega /\omega_L\approx 40$.
This corresponds nicely to $\gamma _{x,\max }\approx 3.2$
(see Fig. 3b) and the relativistic Doppler shift
$\omega /\omega_L=\gamma _{x,\max}^{2}(1+\beta_{x,max})^2\approx 39$.

Apparently, only the velocity component $\beta _{x}$ in the normal direction
of the electron layer and the corresponding $\gamma_{x}=1/\sqrt{1-\beta
_{x}^{2}}=\gamma /\sqrt{1+p_{y}^{2}}$ is relevant for the Doppler shift. The
full shift of $\omega /\omega_L=4\gamma _{\max }^{2}\approx 1400$,
expected for $\gamma _{\max }\approx 19$ (see Fig.3b), is suppressed by the
transverse momentum $p_{y}$ of the electrons in the layer. Of course, in the
present 1D-PIC simulation accounting for three momentum coordinates ($p_{x}$,
$p_{y}$, $p_{z}$), but only one spatial coordinate $x$, scattered
radiation is obtained only in $x$-direction. But also in 2D/3D PIC
simulation, coherently scattered radiation will not be emitted into the
oblique direction of electron motion. This is because of the plane symmetry
of the coherently accelerated electrons and destructive interference of
radiation scattered into directions outside a very narrow cone in normal
direction. Incoherently back-scattered radiation may be observed in other
directions, in particular, when the density of the layer has decreased
strongly, but this incoherent radiation scales with the number scatterers
$N_{sc}$ rather than $N_{sc}^{2}$ for coherent scattering and is therefore
much weaker.

%%%%%%%%%%%%%%%%%%%%%%%%%%%%%%%%%%%%%%%%%%%%%%%
\begin{figure*}[t]
\label{fig7}
\centering\resizebox{0.75\textwidth}{!}{\includegraphics{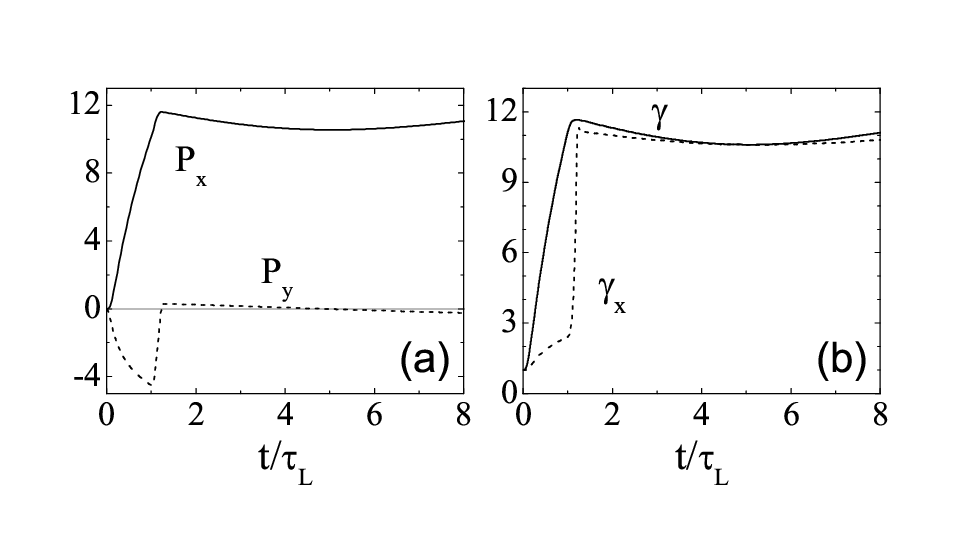}}
\caption{Same as Fig. 5, but for a half-cycle switch pulse. In this
case, the switch pulse turns (a) the transverse momentum
$p_y\rightarrow 0$ and (b) $\gamma_x \rightarrow \gamma \approx 11$
for a longer period of time.}
\end{figure*}
%%%%%%%%%%%%%%%%%%%%%%%%%%%%%%%%%%%%%%%%%%%%%%%%

%%%%%%%%%%%%%%%%%%%%%%%%%%%%%%%%%%%%%%%%%%%%%%%%
\begin{figure*}[t]
\label{fig8}
\centering\resizebox{1.\textwidth}{!}{\includegraphics{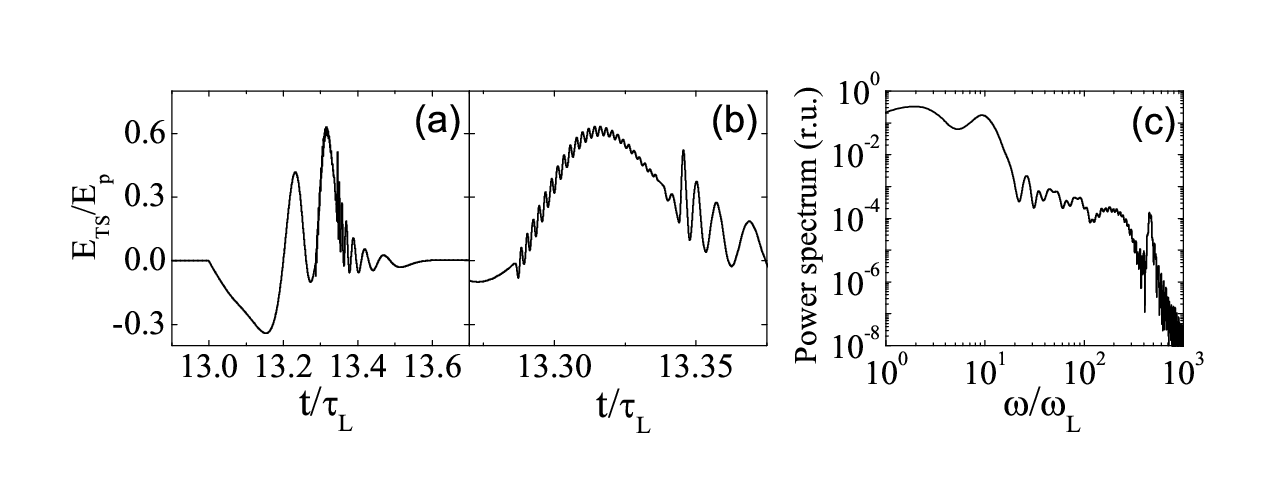}}
\caption{(a,b) Thomson signal and (c) spectrum of probe pulse
corresponding to the case of Fig. 7, but with observer position
moved to $x=13\lambda_L$. The backscattered signal now exhibits fast
oscillations in the time interval $13.25\leq t/\tau_L\leq 13.40$,
shown as zoom in (b). The small amplitude oscillations for
$13.275\leq t/\tau_L\leq 13.335$ correspond to the right edge of the
electron layer (see Fig. 3a). It acts as a relativistic mirrors
moving with $\gamma_x \approx \gamma \approx 11$ and generates the
sharp peak at $\omega/\omega_L \approx 4\gamma^2 \approx 484$ in the
spectrum. The larger amplitude oscillations, seen in Fig. 8b for
$13.335\leq t/\tau_L\leq 13.375$, correspond to the density spike at
the left layer surface. This peak is already decelerating, and this
causes the negative chirp.}
\end{figure*}
%%%%%%%%%%%%%%%%%%%%%%%%%%%%%%%%%%%%%%%%%%%%%%%%%

\subsection{Reflected attosecond pulses after single-cycle switch pulse}

In order to see the full $4\gamma^{2}$ shift, one has to turn the
transverse momentum $p_{y}\rightarrow 0 $ by some additional interaction.
Such interaction can be provided by a short counter-propagating
\textit{switch} pulse. As we have discussed in section 2.3, this pulse
has to have similar strength as the drive pulse and same polarization.
Here we present a simulation in which a single-cycle switch pulse,
having $a_{0}=2.5$ and propagating in $(-x)$-direction,
hits the electron layer at time instant $t/\tau_L=1$,
when the $\gamma$-factor of the electrons is $\approx 11.5$
as seen in Fig. 5b. We choose this early interaction time,
because it is found that the layer rapidly expands in this case and that the
coherently scattered signal would be significantly weaker, had we
waited until $t/\tau_L=5$, when $\gamma _{\max}\approx 19$ is reached.

In Figs. 5a and 5b, the effect of the switch pulse is demonstrated
for electrons located at the right surface of the electron layer (see Fig. 3a).
The transverse momentum makes a sudden excursion to $p_{y}\approx 0$, while the
longitudinal momentum $p_{x}$ is hardly affected. This correlates to a
sudden jump in $\gamma _{x}$ to almost $\gamma _{x}\approx \gamma \approx 11.5$.
We expect that this interaction causes a burst of short wavelength
radiation from the coherently scattered probe pulse. Indeed, we observe
two spikes of attosecond duration in the reflected wave
(see Fig. 6a and zoom in Fig. 6b), a smaller one at $t/\tau_L=7.285$
and a prominent one at $t/\tau_L=7.345$. Correspondingly,
the scattered spectrum (Fig. 6c) exhibits strongly enhanced emission
in the frequency range $\omega /\omega_L\approx 40-2000 $.
Apparently, the two spikes, separated by a time interval
of $\Delta t/\tau_L\approx 0.06$, originate from scattering
at front and rear side of the electron layer (see Fig. 3a).
They are separated by a distance $\Delta x/\lambda_L\approx 0.06$.
Actually, the two signals interfere and produce the conspicuous oscillations
seen in the high-frequency part of the spectrum with
$\Delta \omega /\omega_L\approx 1/0.06\approx 17$.
This interference documents the coherence of two back-scattered signals,
emitted from electron interfaces at different depth in the layer.

Similar to the harmonics spectra arising from reflection from
plasma surfaces \cite{Baeva}, the high-frequency spectrum falls off
according to a power law up to $\omega /\omega_L\approx 4\gamma ^{2}\approx 450$,
consistent with $\gamma _{x}\approx \gamma \approx 11$ at switch time. Also the spectral part
beyond $\omega /\omega_L\approx 450$ is characteristic for attosecond bursts,
showing no clear cut-off. The power law exponent is approximately $-2$, i.e. the fall-off is
smaller than for the value $-8/3$ observed for surface harmonics \cite{Dromey,Baeva}.
We attribute the smaller exponent to the fact that $p_y(t)$ is just touching the $p_y=0$ axis
in Fig. 5a rather than intersecting it, as it does in the case of surface harmonics
(compare Pukhov et al. \cite{Pukhov 2008}).

\subsection{Relativistic mirror after half-wave switch pulse}

Let us finally discuss how a half-cycle switch pulse would change the results.
Again we apply an additional pulse $E=E_s\sin(\tau-\tau_s)$ with $0\leq \tau=t+x \leq \pi$,
the phase $\tau_s$ chosen such that it hits the electron layer at $t/\tau_L\approx 1$,
having same polarization as drive pulse and an amplitude $E_s$ taken such that $p_y\rightarrow 0$
is obtained approximately. The results depicted in Fig. 7a and 7b show that we indeed
find $p_y\approx 0$ and $\gamma_x\rightarrow \gamma \approx 11$ for $t/\tau_L > 1$ when
selecting the appropriate parameters. Probing this relativistic mirror as before
by a perpendicularly polarized probe pulse, we find the back-scattered signal of Fig. 8a
with the spectrum of Fig. 8c. On top of the reflected signal without switch pulse
(compare Fig. 4a), we observe a high-frequency feature (zoomed in Fig. 8b),
now showing many fast oscillations after the switch pulse has turned the electron momentum
into normal direction. The zoom in Fig. 8b exhibits two regions,
low-amplitude fast oscillations for $13.285 \le t/\tau_L \le 13.345$ corresponding to
the front interface of the electron sheet and high-amplitude chirped oscillations
corresponding to the prominent density spike at the rear side of the layer,
which is already decelerating. Since the probe pulse now sees the full
$\gamma \approx 11$ over an extended time,
the reflected spectrum exhibits a clear peak
at $\omega/\omega_L\approx 4\gamma^2 \approx 480$.

\section{Conclusions}

In this paper, we have explored coherent Thomson back-scattering from
relativistic electron layers, blown out from ultra-thin solid foils by
high-contrast few-cycle laser pulses. For this to happen, the laser
amplitude $a_{0}$ has to be larger than the electrostatic field
$\varepsilon_{0}$ building up when separating all electrons from ions in the
foil. It requires ultra-thin foils: already a mono-atomic carbon layer
(graphene, used here as reference material) produces $\varepsilon _{0}\simeq
1$. A second requirement is ultra-high contrast of the laser pulses to
avoid premature foil destruction. Though demanding, these conditions may be
met in near-future experiments.

Here we have investigated the generation of relativistic electron layers by
schematic single-cycle laser pulses with $a_{0}=5,$ making use of well known
analytical theory and 1D/2D PIC simulation. The ultra-thin foils are
transparent to light, and electron acceleration occurs by direct action of
the laser fields, admitting an approximate analytical description in terms of
single-electrons in a plane light wave. Equivalent electrons then move in
planes, and coherent backscattering of probe radiation occurs according to
normal mirror rules due to phase selection. Conditions for coherent
backscattering are discussed in a companion paper (paper II ),
and the scaling of reflectivity with foil density, thickness, and $\gamma $-factor
are derived there \cite{Wu}.

An important point is that, due to the laws of direct laser acceleration,
high-$\gamma $ electrons necessarily have high transverse momentum $p_{y}$
and move under an angle relative to the direction of laser propagation,
which is also the normal direction of the electron layer. The relativistic
Doppler shift of reflected light is therefore reduced to
$\omega /\omega_L\approx 4\gamma_{x}^{2}=4\gamma ^{2}/(1+p_{y}^{2})$,
since only $\gamma_{x}=1/\sqrt{1-\beta _{x}^{2}}$ related
to normal $x$-direction is relevant.
Methods to switch $p_{y}\rightarrow 0$ by an additional counter-propagating
laser pulse or by a static magnetic field have been discussed. Various cases
are demonstrated by 1D-PIC simulation.

Probe radiation backscattered from a driven electron layer typically shows a
reflected signal with a broad chirped spectrum and a cut-off at $\omega
/\omega_L\approx 4\gamma _{x}^{2}$. The chirp is due to layer acceleration
and may well extend over several octaves. Disrupting the smooth layer motion
by a counter-propagating single-cycle switch pulse changes the spectrum
of the back-scattered probe radiation significantly.
Having an appropriate amplitude, the switch pulse turns
$p_{y}\rightarrow 0$ for a short moment, which is
much shorter than a laser cycle. The probe radiation then
experiences the full $4\gamma ^{2}$ Doppler shift for this short moment,
and a single attosecond burst of high-frequency radiation is backscattered.
A broad spectrum in the region of $\omega /\omega_L \approx 4\gamma ^{2}$
is observed, which looks quite similar to the surface harmonics,
recently observed experimentally \cite{Dromey}. We have also investigated
the more speculative case of a unipolar half-wave switch pulse
which may turn $p_{y}\rightarrow 0$ for a longer time interval.
The PIC simulation then shows an extended period of full $4\gamma ^{2}$ Doppler shift
and, correspondingly, a strong peak at  $\omega/\omega_L \approx 4\gamma ^{2}$
in the spectrum.

Though the long term goal is to use the scattered probe radiation for
flexible generation of controlled attosecond VUV- and X-ray pulses,
the immediate application will be to analyze the electron layer and
its angular and temporal evolution in all detail. Already the few cases
shown in this paper illustrate the abundance of detailed information
available from the spectra. The present study is only a first step.
Many questions arise concerning what happens
with more realistic few-cycle pulses than those used here.
Also mono-atomic foils may not be available immediately,
and acceleration of foils a few nanometer thick needs to be studied.
This is planned for subsequent investigations.

\section{Acknowledgements}
The authors would like to thank D. Habs, F. Krausz, A. Pukhov, S. Rykovanov, and L. Veisz
for helpful discussions concerning this paper. H.C. Wu thanks the Alexander
von Humboldt Foundation for a scholarship. This work was supported in part by
the DFG project Transregio TR18, by the Munich Centre for Advanced Photonics (MAP),
and by the Association EURATOM - Max-Planck-Institute for Plasma Physics.

\end{document}